\documentclass[amsmath,amssymb,aps,prd,reprint,nofootinbib]{revtex4-2}
\usepackage[utf8]{inputenc}
\usepackage{graphicx}
\usepackage{dcolumn}
\usepackage{bm}
\usepackage{feynmf}
\usepackage{color}
\usepackage{ulem}

\newcommand{\be}{\begin{equation*}}
\newcommand{\ee}{\end{equation*}}
\newcommand{\ba}{\begin{eqnarray*}}
\newcommand{\ea}{\end{eqnarray*}}
\newcommand{\ban}{\begin{eqnarray}}
\newcommand{\ean}{\end{eqnarray}}

\newcommand{\bw}{\begin{widetext}}
\newcommand{\ew}{\end{widetext}}

\newcommand{\pppp}{\ensuremath{\mathord{+}\mathord{+}\mathord{+}\mathord{+}}}
\newcommand{\mmmm}{\ensuremath{\mathord{-}\mathord{-}\mathord{-}\mathord{-}}}
\newcommand{\pmpm}{\ensuremath{\mathord{+}\mathord{-}\mathord{+}\mathord{-}}}

\newcommand{\Imag}{\mathop{\mathrm{Im}}}

\begin{document}

\title{Seeking for resonances in unitarized one-loop graviton-graviton scattering} 

\author{Rafael L. Delgado}
\email{rafael.delgado@upm.es}
\affiliation{Matemática Aplicada a las TIC, ETSIS de Telecomunicación,\\
Universidad Politécnica de Madrid (Campus Sur) 28031 Madrid, Spain}

\author{Antonio Dobado}
\email{dobado@fis.ucm.es}
\affiliation{Departamento de Física Teórica and IPARCOS \\
Universidad Complutense de Madrid, 28040 Madrid, Spain}

\author{Domènec Espriu}
\email{espriu@icc.ub.edu}
\affiliation{Departament de Física Quàntica i Astrofísica and Institut de Ciències del Cosmos (ICCUB) \\
Universitat de Barcelona, 08028 Barcelona, Catalonia, Spain}
\date{\today}

\begin{abstract}
Some effective field theories exhibit dynamical resonances that, when properly included, mitigate their bad behavior at high energies. Unitarization of the partial wave amplitudes is the preferred method to unveil such resonances. Interpreting the Einstein-Hilbert theory in the spirit of effective Lagrangians, we implement the Inverse Amplitude Method and unitarize the one-loop level graviton-graviton scattering in pure gravity. Due to the presence of infrared divergences, the analysis requires a careful treatment of the infrared region and the introduction of infrared regulators, carefully selected in order to fulfill perturbative unitarity. 
\end{abstract}

\maketitle

\section{Introduction}
The Einstein-Hilbert (EH) Lagrangian and the effective chiral 
Lagrangian~\cite{EChL}, quite familiar to low-energy QCD practitioners, share a number of common characteristics. Like the
effective chiral Lagrangian, EH is also a non-renormalizable theory. It is also
described, considering the most relevant term, by a dimension two operator, containing in
both cases, two derivatives of the dynamical variable. Both Lagrangians contain necessarily a
dimensionful constant in four dimensions; the counterpart of $f_\pi$ in the pion Lagrangian is the
Planck mass $M_P$. Both theories are non-linear and, finally, both describe the interactions of
massless quanta. There is a fundamental difference, however, because the 
theory described by EH, gravity, is a gauge theory thus fixing quite rigidly its structure (but allowing in principle for higher dimensional operators also gauge invariant).

The analogy becomes particularly clear when one linearizes gravity around a given background, such as e.g. Minkowski space-time. There have been in the past a number of theoretical developments considering such as expansion in the spirit of effective field theories. It is appropriate here to quote the work of Donoghue and others \cite{Donoghueetal}. Taking into account the usual normalization
of the EH action
\begin{equation}
{\cal L}= \frac{M_P^2} {16\pi}\sqrt{-g} {\mathcal R} + \ldots \qquad M_P^2 = \frac{1}{G},
\end{equation}
$G$ being Newton's constant and the dots standing for higher dimensional counterterms that are required to absorb
order by order the divergences appearing in perturbation theory (or other possible contributions from short distance physics). We can expand the metric as:
\begin{eqnarray}
g_{\mu \nu} & \equiv & \eta_{\mu \nu} + \kappa h_{\mu \nu} \,
 \\
g^{\mu \nu} &=& \eta^{\mu \nu} - \kappa h^{\mu \nu} + \kappa^2 h^{\mu
\lambda} h_{\lambda}^{~\nu} + \ldots
\end{eqnarray}
and the scalar curvature:
\begin{equation}
{\mathcal R} = \kappa \left[ \Box h^{\lambda} _{~\lambda} - \partial_{\mu}
\partial_{\nu} h^{\mu \nu} \right] + {\cal O} (h^2),
\end{equation}
where 
\begin{equation}
\kappa^2 \equiv \frac{32\pi}{M_P^2}= 32 \pi G.
\end{equation}
Indices are raised and lowered with $\eta_{\mu \nu}$.

In pion physics the effective Lagrangian is:
\begin{equation}
{\cal L}= \frac{f_\pi^2}{4} {\rm Tr\,} \partial_\mu U \partial^\mu U^\dagger
 + \ldots
\end{equation}
Here we assume $SU(2)_L\times SU(2)_R$ global chiral symmetry.
Again, the dots denote quantum counterterms and the higher order operators can contain
contributions from short distance physics.
One writes
\begin{equation}
U= I + i \frac{\tilde\pi}{f_\pi}+ . . . 
\end{equation}
where $\tilde \pi = \pi^a \tau^a$ and Cartan normalization is assumed for the $SU(2)$ 
generators. Therefore $\kappa$ plays the same role as $f_\pi^{-1}$ (up to a factor $\sqrt{2}$ in our normalizations). Quantum corrections in gravity are 
analogous to the weak field expansion in pion physics.

If we consider both the pion Lagrangian and the EH Lagrangian as effective theories we can, at least naively, attribute to each one a range of validity based on power counting. The corresponding unitarity cut-offs would be of order $4\pi f_\pi$ and $M_P$, respectively.

In EH theory quantum corrections are notoriously difficult to compute.
Graviton-graviton scattering amplitudes at the tree level are given in~\cite{tree}, while
one-loop corrections were first carried out by Hooft and Veltman~\cite{tHV} who found that the 
counterterms needed to deal with the ultraviolet (UV) divergences were proportional to ${\mathcal R}^2$ and ${\mathcal R}_{\mu\nu}{\mathcal R}^{\mu\nu}$.
On-shell, i.e. using the lowest-order equations of motion these counterterms vanish. Therefore one-loop $S$ matrix elements in pure EH gravity are free from UV divergences. Thus the only possible divergences are infrared (IR), as it is the case of the elastic graviton scattering considered in this paper. A general study of the structure of the IR
divergences in EH gravity, considered as a quantum field theory, can be found in the seminal paper by Weinberg \cite{weinb}.

Computation of the next-to-next-to-leading order in pure gravity was done by Goroff and Sagnotti  
\cite{Goroff:1985th} finding a net counterterm of the form
${\mathcal R}_{\alpha\beta}^{\;\;\;\;\gamma\delta}
{\mathcal R}_{\gamma\delta}^{\;\;\;\;\rho\sigma}{\mathcal R}_{\rho\sigma}^ {\;\;\;\;\alpha\beta}$.
This counterterm will not play any role in the subsequent discussion.

Taking the analogy between pion physics and quantum gravity at face value, immediately comes to mind the following issue: we know that unitarization of pion scattering amplitudes in the context of low-energy hadron physics leads to poles that restore the unitary behaviour that is lacking in the above
pion chiral Lagrangian. These poles correspond to physical (albeit unstable) particles or dynamical resonances such as
the $\rho$ mesons or the $\sigma$ particle. These states correspond to poles in the second Riemann sheet of the amplitude when this is extended to the complex plane. The real part should of course stay below the theory cut-off at $4\pi f_\pi$. Could it be that such a phenomenon occurs in 
the EH theory? If so this would lead at the very least to some dynamical resonances that would hint to the presence -as it happens in pion physics- of more fundamental degrees 
of freedom.

In order to answer this question in a way that is similar to the pion physics techniques we need two ingredients: a computation of the one-loop graviton-graviton scattering amplitude, done by Dunbar and Norridge using string theory methods \cite{Dunbar:1994bn} and the tree-level contribution from higher dimensional operators. At the next-to-leading order the latter are absent on-shell as we mentioned previously. It is possibly worth mentioning that the fact that ${\mathcal O}(p^4)$ are absent does not necessarily preclude the possibility of resonances being present. Recent studies in the context of the
strongly interacting symmetry breaking sector of the Standard Model show that ${\mathcal O}(p^2)$ may suffice to produce such singularities \cite{dobetal}.

In \cite{Blas:2020och} the authors use a simple unitarization method relying only on the tree-level graviton scattering amplitude (see Appendix I) and they claim that a scalar graviton-graviton resonance with quantum numbers $0^{++}$ can be identified, well below the Planck mass and with a sizable width. As it will be discussed below, our conclusions differ significantly from those derived in \cite{Blas:2020och}. Our results are based on a careful consideration of the IR  divergences that are regulated in a way that preserves perturbative unitarity at the one-loop level. We then apply the inverse amplitude method (IAM) to unitarize the amplitudes and study their singular points. Although the resulting partial waves do exhibit a surprisingly rich structure, no resonances physically acceptable seem to survive 
to be considered as new states, at least in pure EH gravity.

\section{One-loop scattering amplitudes}

In this work we are considering the elastic scattering of two gravitons with initial momenta and helicities $p_1, \lambda_1$
and $p_2, \lambda_2$ to $p_3, \lambda_3$
and $p_4, \lambda_4$ in the final state. The corresponding helicity amplitude is defined as:
\begin{equation}
\nonumber
T_{\lambda_1\lambda_2\lambda_3\lambda_4}(s,t,u)=\left<p_3,\lambda_3;p_4,\lambda_4\mid T \mid p_1,\lambda_1;p_2,\lambda_2 \right>,
\end{equation}
where $s=(p_1+p_2)^2, t=(p_1-p_3)^2$ and $u=(p_1-p_4)^2$. $T$ is the standard reaction matrix related to the $S$ matrix by: $S= I + i(2\pi)^4\delta^{(4)}(P_f-P_i)T$, with $P_i=p_1+p_2$ and $P_f= p_3+p_4$.
The helicities $\lambda_i$ can only take the values $+2$ or $-2$ which by simplicity will be denoted $\lambda_i=+,-$  respectively. By using $P$ and $T$ invariance and crossing it is possible to relate different helicity amplitudes in such a way that one gets just three independent functions:
\begin{eqnarray}
A(s,t,u)=T_{++++}(s,t,u)  \nonumber \\
B(s,t,u)=T_{+++-}(s,t,u)   \nonumber \\
C(s,t,u)=T_{++--}(s,t,u).   \nonumber
\end{eqnarray}
Expanding these functions according to the number of loops one has for example:
\begin{equation}
A = A^{(0)} + A^{(1)}+\dots
\end{equation}
At the tree level (no loops) the result is very simple, in spite of the 
complexities of the Feynman diagrams involved:
\begin{eqnarray}
A^{(0)}(s,t,u) & = & \frac{8\pi}{M_P^2}\frac{s^3}{tu}  \nonumber \\
B^{(0)}(s,t,u) & = & 0  \nonumber \\
C^{(0)}(s,t,u) & = & 0, \nonumber
\end{eqnarray}
where $M_P$ is again the Planck mass. Notice that at this level, the amplitudes are order $p^2/M_P^2$. Since gravitons are massless we have two poles corresponding to the $t$ and $u$ channels infrared virtual gravitons, as expected.

At the one-loop level, using dimensional regularization with $D=4-2\epsilon$, we have for small $\epsilon$ \cite{Dunbar:1994bn}:
\begin{equation}
 A^{(1)}(s,t,u) = 8\frac{s^4}{M_P^4}\large[...\large]
\end{equation}
where:
\begin{equation}
 \large[...\large]=(N'_\epsilon + \log \nu^2)(...) + \{...\}+\frac{f(t/s,u/s)}{2s^2}    \nonumber
\end{equation}
with:
\begin{equation}
 N'_\epsilon= \frac{1}{\epsilon}+\log (4 \pi)-\gamma
\end{equation}
and $\nu$ is an, in principle, arbitrary energy scale. Also
\begin{equation}
(...)=\frac{s \log(-s)+t \log(-t)+u \log(-u)}{stu}    \nonumber
\end{equation}
and
\begin{multline}
\{...\}=\frac{1}{stu} \Big[ s\log(-t)\log(-u) + t\log(-u)\log(-s) \\
 + u\log(-s)\log(-t)\Big],    \nonumber
\end{multline}
and finally the function $f(t/s,u/s)$ is given by:
\begin{widetext}
\begin{multline}
f(t/s,u/s) =\frac{1}{s^6} 
(t+2u)(u+2t)(2t^4+2t^3u-t^2u^2+2tu^3+2u^4)\left(\log^2\frac{t}{u}+\pi^2\right) \\  
+ \frac{1}{30s^5}(t-u)(341t^4+1609t^3u+2566t^2u^2+1609tu^3+341u^4)\log\frac{t}{u} \\
+ \frac{1}{180s^4}(1922t^4+9143t^3u+14622t^2u^2+9143tu^3+1922u^4). \nonumber 
\end{multline}
\end{widetext}
Notice that here we have introduced a new energy scale $\nu$ not present in the original Dunbar and Norridge result \cite{Dunbar:1994bn}
in order to make the one-loop amplitude dimensionally consistent. In the following and, at the end of section IV, we will clarify the role played by this scale.   

Introducing the new energy scale $\nu$ into the logarithms in  the expression above we define:
\begin{equation}
\{...\}_\nu=\frac{s \log\left(-t/\nu^2\right)\log\left(-u/\nu^2\right)+\dots
}{stu}.    \nonumber
\end{equation}
Similarly one could introduce $(...)_\nu$, but in this case:
\begin{equation}
(...)_\nu=(...)
\end{equation}
since $s+t+u=0$ on shell. Then it turns out that
\begin{equation}
    \{...\}_\nu=\log \nu^2 (...) +\{...\}.
\end{equation}
Therefore we have:
\begin{equation}
 A^{(1)}(s,t,u) = 8\frac{s^4}{M_P^4}\left[N'_\epsilon(...)+
 \{...\}_\nu+\frac{f(t/s,u/s)}{2s^2}\right] . \nonumber
\end{equation}
Next we define a new energy scale $\mu$ as:
\begin{equation}
 \log \frac{\mu^2}{\nu^2} =N'_\epsilon  \nonumber.
\end{equation}
The IR limit $\epsilon \rightarrow 0$ can be taken as $\nu \rightarrow 0$ while keeping the new scale $\mu $ fixed. 
Then we have the IR finite result:
\begin{multline}
A^{(1)}(s,t,u) = 8\frac{s^4}{M_P^4}\left[\frac{s \log(\frac{-t}{\mu^2})\log(\frac{-u}{\mu^2})+\dots}{stu}
\right.\\\left.
+ \frac{f(t/s,u/s)}{2s^2}\right], \nonumber
\end{multline}
where the meaning of the finite scale $\mu$ will be clarified below. A different 
choice as for example:
\begin{equation}
 \log \frac{\mu^2}{\nu^2}  =\frac{1}{\epsilon} \nonumber,
\end{equation}
amounts just to the same result but trading $\mu$ by $\mu'$ where:
$\log(\mu'^2/\mu^2)=\log(4\pi)-\gamma$.

In any case, it is very important to realize that, the introduction of the $\log \nu^2$ term into the one-loop 
amplitude, renders this amplitude IR finite to the cost of introducing a new scale $\mu$. The precise meaning of this
finite new scale will be clarified at the end of section IV by comparison with other computations.

It is also very important to stress that, as discussed in the introduction, one-loop matrix elements are UV finite in EH pure gravity. Therefore, the $\epsilon$ pole found in the one-loop elastic graviton scattering is purely IR and it is entirely produced by low-energy massless gravitons. A detailed study of the IR divergences appearing in the one-loop graviton-graviton elastic amplitudes can be found in \cite{dt} and \cite{Dunbar:1995ed}.
In this work we deal with the IR divergencies just by introducing
the new scale $\nu$, playing the role of IR cut-off, which requires the introduction of a new finite scale $\mu$.

On the physical region $s=E_{CM}^2+i0$, with $E_{CM}$ being the total center of mass energy, and then
$\log (-s)= \log s - i \pi$. Therefore:
\begin{equation}
 \Imag A^{(1)}(s,t,u)= - \frac{8 \pi s^2}{M_P^4}\left(\frac{1}{t}\log \frac{-t}{\mu^2}+ \frac{1}{u}\log\frac{-u}{\mu^2}\right).
\end{equation}
The other two relevant one-loop functions are:
\begin{eqnarray}
 B^{(1)}(s,t,u) & = &   \frac{s^2+t^2+u^2}{90 M_P^4}\nonumber,   \\
  C^{(1)}(s,t,u) & = &  - \frac{s^2+t^2+u^2}{30 M_P^4}\nonumber.
\end{eqnarray}
These functions are much simpler than $A^{(1)}(s,t,u)$ being real and IR finite and will not be considered in the following.
Notice however that all the three one-loop functions are of the order of $p^4/M_P^4$ as expected. Therefore loop expansion is
a low-energy expansion valid for energies small compared with the Planck mass $M_P$.

\section{Partial waves and elastic unitarity}\label{sec:partialwave}
For well behaved helicity amplitudes the partial waves are defined in principle as:
\begin{equation}
a_{J\lambda_1,\lambda_2,\lambda_3\lambda_4}(s)=\frac{1}{64\pi}\int_{-1}^{1}d(\cos \theta)   d^J_{\lambda,\lambda'}(\theta) T_{\lambda_1\lambda_2\lambda_3\lambda_4} (s,\theta)\nonumber
\end{equation}
where $\lambda=\lambda_1 - \lambda_2$,   $\lambda'=\lambda_3 - \lambda_4$ and we have used $t=-(s/2)(1-x)$ 
and $u=-(s/2)(1+x)$ with $x=\cos \theta$. When these integrals are well defined for any $J$ we have:
\begin{equation}
T_{\lambda_1\lambda_2\lambda_3\lambda_4} (s,\theta)=32 \pi\sum_J [J]
d^J_{\lambda,\lambda'}(\theta)a_{J\lambda_1\lambda_2\lambda_3\lambda_4}(s), \nonumber
\end{equation}
where $[J]=(2J+1)$. 

For physical $s$ ($ s =  E_{CM}^2 +i0 $) elastic (two-particle states) unitarity reads:
\begin{eqnarray}
\Imag T_{\lambda_1\lambda_2\lambda_3\lambda_4} (s,\theta) & = & \nonumber\\
\frac{1}{128 \pi^2}\sum_{\lambda_a \lambda_b}
\int d \Omega' 
& T_{\lambda_1\lambda_2\lambda_a\lambda_b} (s,\theta')
 T_{\lambda_a\lambda_b\lambda_3\lambda_4}^* (s,\theta'')
\nonumber
\end{eqnarray}
where $\theta'$ and $\theta''$ are the scattering angles 
between the initial state and the intermediate state and the 
scattering angle between the intermediate state and the final state respectively. This equation can be written in terms 
of the partial waves as:
\begin{equation}
\Imag a_{J\lambda_1\lambda_2\lambda_3\lambda_4} (s)  =  
\sum_{\lambda_a \lambda_b}
 a_{J\lambda_1\lambda_2\lambda_a\lambda_b} (s)
 a_{J\lambda_a\lambda_b\lambda_3\lambda_4}^* (s).
\nonumber
\end{equation}
On the other hand the amplitude loop expansion can be translated into the partial waves:
\begin{equation}
a_{J\lambda_1\lambda_2\lambda_3\lambda_4} (s)  =  
a_{J\lambda_1\lambda_2\lambda_3\lambda_4}^{(0)} (s)+ 
 a_{J\lambda_1\lambda_2\lambda_3\lambda_4}^{(1)} (s)+\dots
\nonumber
\end{equation}
Then, the lowest order perturbative unitarity relation becomes:
\begin{equation}
\Imag a^{(1)}_{J\lambda_1\lambda_2\lambda_3\lambda_4} (s)  =  
\sum_{\lambda_a \lambda_b}
 a^{(0)}_{J\lambda_1\lambda_2\lambda_a\lambda_b} (s)
 a_{J\lambda_a\lambda_b\lambda_3\lambda_4}^{(0)*} (s).
\nonumber
\end{equation}
However, all this formalism cannot be applied directly to elastic graviton scattering because of the presence of IR divergences.
In particular the partial waves are ill defined because of the behavior of the helicity amplitudes for $x=\cos\theta$
close to 1 and -1. One possible way to deal with this problem is by introducing the regularized amplitudes 
$\tilde T_{\lambda_1\lambda_2\lambda_3\lambda_4}^{\eta} (s,\theta)$ defined as:
\begin{equation}
\tilde T_{\lambda_1\lambda_2\lambda_3\lambda_4}^{\eta}(s,\theta)= T_{\lambda_1\lambda_2\lambda_3\lambda_4} (s,\theta)
\end{equation}
for $x \in[-1+\eta,1-\eta]$ with $0<\eta < 1$ and
\begin{equation}
\tilde T_{\lambda_1\lambda_2\lambda_3\lambda_4}^{\eta} (s,\theta)= 0
\end{equation}
otherwise. In this way $\tilde T_{\lambda_1\lambda_2\lambda_3\lambda_4}^{\eta} (s,\theta)$ is a bounded
function of $x$ with only two discontinuity points at $x= -1+\eta$ and $x= 1-\eta$ and then we can apply
the partial wave formalism described above to it. In particular we can define the partial waves as:
\begin{multline}
a_{J\lambda_1\lambda_2\lambda_3\lambda_4}(s,\eta) = \\ \frac{1}{64\pi}\int_{-1}^{1}d(\cos\theta) d^J_{\lambda,\lambda'}(\theta) \tilde  T_{\lambda_1\lambda_2\lambda_3\lambda_4}^{\eta} (s,\theta) = \\
\frac{1}{64\pi}\int_{-1+\eta}^{1-\eta}d(\cos \theta)   d^J_{\lambda,\lambda'}(\theta)  T_{\lambda_1\lambda_2\lambda_3\lambda_4} (s,\theta)
\end{multline}
which are IR finite. From these partial waves we can recover the regularized amplitude as:
\begin{equation}
\tilde T_{\lambda_1\lambda_2\lambda_3\lambda_4}^\eta (s,\theta)=32 \pi\sum_J [J]
d^J_{\lambda,\lambda'}(\theta)a_{J\lambda_1\lambda_2\lambda_3\lambda_4}(s,\eta). \nonumber
\end{equation}
In the following we will study the meaning of these regularized amplitudes and partial waves.

\section{Perturbative unitarity}\label{sec:pertunit}
By using the above definitions it is very easy to compute the lowest order contribution to different regularized helicity amplitudes partial waves. For example for $J \le 4$ one has:
\ba \label{(pw0)}
a_{0++++}^{(0)}(s,\eta) &=& \frac{s}{2M_P^2}\log\frac{2}{\eta} +O(\eta)\\
a_{2++++}^{(0)}(s,\eta) &=& \frac{s}{2M_P^2}
  \left(\log\frac{2}{\eta} - 3\right) +O(\eta)\\
a_{4++++}^{(0)}(s,\eta) &=& \frac{s}{2M_P^2}
  \left(\log\frac{2}{\eta} - \frac{25}{6} \right)+O(\eta) \\
a_{0+-+-}^{(0)}(s,\eta) &=& a_{2+-+-}^{(0)}(s,\eta) = 0 \\
a_{4+-+-}^{(0)} (s,\eta)&=& \frac{s}{4M_P^2}
  \left(\log\frac{2}{\eta} - \frac{363}{140} \right)+O(\eta) \\
a_{J+--+}^{(0)}(s,\eta) &=& a_{J+-+-}^{(0)}(s,\eta), \\
\ea
where we are showing the results modulo $O(\eta)$ corrections, i.e. only the contributions dominant in the asymptotic regime $\eta \ll 1$. However, it is very important to stress that, in order to have a proper reconstruction of the full helicity amplitude, one needs to use  the partial waves with the exact $\eta$ dependence, and not only the part dominant for small $\eta$. More specifically the amplitude obtained summing the different $a_{J\lambda_1\lambda_2\lambda_3\lambda_4}(s,\eta\ll 1)$ contributions does not converge to $\tilde T_{s,\eta\ll 1}$ because the limit $\eta \ll 1$ and the sum $\sum_J$ do not commute.

With the modified amplitude $\tilde T^{\eta}(s,\theta)$ one would expect an elastic scattering unitarity relation like:
\begin{eqnarray}
\Imag \tilde T_{\lambda_1\lambda_2\lambda_3\lambda_4}^\eta (s,\theta) & = & \nonumber\\
\frac{1}{128 \pi^2}\sum_{\lambda_a \lambda_b}
\int_R d \Omega' 
& \tilde T_{\lambda_1\lambda_2\lambda_a\lambda_b}^\eta (s,\theta')
\tilde T_{\lambda_a\lambda_b\lambda_3\lambda_4}^{\eta*} (s,\theta'')
\nonumber
\end{eqnarray}
where $R$ represents the two body phase space region  allowing  only states where $s=4E_{CM}^2>  \mu^2$, i.e. the scale regulator introduced above for the one-loop contribution to the amplitude. Clearly, this regulator must be related in some way with the $\eta$ parameter introduced in the modified amplitude. In addition, crossing requires also $t <-\mu^2$ and the same for $t'$ and $t''$. Now by trading $x=\cos \theta $ by $t$ we find:
\begin{equation}
    \int^{1-\eta}_{-1+\eta}dx= \frac{s}{2}\int^{t_{max}}_{t_{min}}dt=\frac{s}{2}\int^{-\mu^2}_{-s+\mu^2}dt
\end{equation}
which leads to the simple relation:
\begin{equation}
    \mu^2 = \eta \frac{s}{2}.
\end{equation}
Therefore, by using this equation we are assuming that the scale $\mu$ is defining the difference between soft and hard gravitons, i.e. the ones that can be effectively detected. In order to check this equation we can try to see how it works with the unitarity relations in terms of the partial waves. For the sake of simplicity, and also because it is the most interesting process for us 
in this work, we will concentrate in the particular case described by the amplitude $T_{++++}$. Also, to relieve the notation we will omit the subscripts \pppp\ in the following. Then we can compute the exact
(i.e. to all orders in $\eta$) partial wave:
\be
a_{0}^{(0)}(s,\eta) = \frac{s}{2M_P^2}\log\frac{2-\eta}{\eta} \nonumber.
\ee
On the other hand we have:
\be
\Imag a_{0}^{(1)}(s,\eta) = \frac{s^2}{4M_P^4}\log\frac{2-\eta}{\eta}\left[\log\frac{\eta(2-\eta)}{4}+2 \log\frac{s}{\mu^2}\right] \nonumber.
\ee
Thus, it is clear that defining $\eta$ as $\eta=2 \mu^2/s$ we have:
\be
\Imag a_{0}^{(1)}(s,\eta) = \left(a_{0}^{(0)}(s,\eta)\right)^2 \nonumber
\ee
to all orders in $\eta$. However this is much more that one could have expected. Because of the way in which the $\mu$ scale was introduced one
should expect the relation $\eta=2\mu^2/s$ to work only for small enough $\mu$. This is because this scale was introduced in terms of $\epsilon$, which is supposed to be a small quantity in the dimensional regularized one-loop amplitude considered.  
In fact, for $J\ne 0$ this relation works only up to order $\eta$ i.e.:
\be
\Imag a_{J}^{(1)} = \left(a_{J}^{(0)}\right)^2 + O(\eta). \nonumber
\ee
Even more, for the case \pmpm\ the corresponding equation is only fulfilled up to constant terms and therefore only applies for 
the divergent terms in the limit $\eta \rightarrow 0$.

In the following we will focus on the \pppp\ and \mmmm\ partial waves $\eta$-regularized and where we have applied the substitution $\eta$ by $2 \mu^2/s$. Then, the general form of the leading order (LO) partial waves is:
\begin{equation}  \label{mumeaning}
  a_J^{(0)}(s) = \frac{s}{2M_P^2}\left(\log\frac{s}{\mu^2}-b_J\right),  
\end{equation}    
 where $b_0=0, b_2=3, b_4=25/6,\dots$ and, at the one-loop level:
 \begin{multline}   
 a_J^{(1)}(s) = \frac{s^2}{4\pi M_P^4}\left[  \left(\log\frac{s}{\mu^2}-b_J\right)^2F\left(\frac{s}{\mu^2}\right)\right.\\
   \left.+c_J\log\frac{s}{\mu^2}+d_J \right]\end{multline}
where both equations have to be understood up to order $O(\mu^2/s)$. The first $c_J$ and $d_J$ constants can be obtained from eq.~\ref{(pw1)} of Appendix II and 
the $F(s/\mu^2)$ function is defined as:
\begin{equation}   
F\left(\frac{s}{\mu^2}\right)=\log\frac{s}{\mu^2}-\log\frac{-s}{\mu^2}
\end{equation}
and it equals $i\pi$ for $\Imag s \ge 0$ and $-i\pi$ for $\Imag s < 0$.
Therefore, our $\eta$ IR regularized partial waves show, at next to leading order (NLO), the expected unitarity right cut (RC) along the positive real axis, and the expected left cut (LC) along the negative real axis thus consistent with the appropriate analytical behavior. However, these partial waves are unitary only perturbatively. This in particular implies that at higher energies ($s \simeq M_P^2$) we will find strong violations of (elastic) unitarity.

Now, in order to clarify a little more the meaning of the scale $\mu$ in our equations we can proceed in a similar way as in \cite{Blas:2020och} as follows. First we consider the IR regularized $\bar S^J$ matrix
\begin{equation}
\bar S^J = 1 + 2 i \bar  a_J
\end{equation}
where the $\bar  a_J$ lowest order is given by:
\begin{equation}
\bar  a_J^{(0)}(s) = \frac{s}{2M_P^2}\left(\log\frac{s}{\nu^2}-b_J\right),  
\end{equation}    
with $\nu$ being an IR cutoff. Next one introduces a new $S^J$ matrix as $S^J=S_c^{-1}\bar S^J$ with $S_c$ being the Weinberg phase \cite{weinb} given in this case by:
\begin{equation}
S_c= \exp \left(-i \frac{s}{M_P^2}\log\frac{\nu^2}{{\cal L}^2}\right) .
\end{equation}
Here again $\nu$ is an IR cutoff and ${\cal L}$ is a scale separating soft from hard gravitons. 
Thus one gets:
\begin{equation}
 S^J = 1 + 2 i  a_J=S_c^{-1}( 1 + 2 i \bar  a_J)
\end{equation}
with
\begin{equation}
 S^J = 1 + 2 i  a_J^{(0)} + O(s^2/M_P^4)  
\end{equation}
and
\begin{equation}
a_J^{(0)}(s)=\frac{s}{2M_P^2}\left(\log\frac{s}{{\cal L}^2}-b_J\right)
\end{equation}
which is trivially well defined and finite in the IR limit $\nu \rightarrow 0$ since the $\log \nu^2$ terms cancel in the final result. 
Now, by comparison of this equation with eq.~\ref{mumeaning} one arrives to the identification 
$\mu={\cal L}$, i.e. our scale $\mu$ must be understood as the scale used to define soft and hard gravitons
as indeed was suggested above. The same conclusion is obtained by expanding $S_c$ up to order $s^2/M_P^4$
and considering the perturbative unitarity relation $\Imag a_{J}^{(1)} = \left(a_{J}^{(0)}\right)^2  $.

In \cite{dt} Donoghue a Torma showed explicitly that the one-loop graviton scattering differential cross-section is free of IR divergences. For that they add to the IR divergences free tree level result two contributions: the one loop result, with the $1/\epsilon$ singularity and a $\log s$ term, and the soft graviton emitting tree level amplitude integrating the phase space of the additional soft graviton up to momentum
that we can identify with the soft-hard separation scale
${\cal L}$ previously defined,
thus producing another $1/\epsilon$ singularity term with opposite sign and a $\log {\cal L}$ term. In the resulting one-loop differential cross-section the $1/\epsilon$ terms exactly cancel and the final result depends on $\log (s/{\cal L}^2)$. By comparison with our results in this work, it is clear that the introduction of the term $\log \nu^2$ in the one-loop amplitude used here produces a similar effect on the the IR finite result as including the soft graviton emission contribution and that our finite scale $\mu$ plays the same role as the scale that separates hard from soft gravitons in \cite{dt}.

Therefore, by comparison with the results in \cite{Blas:2020och} and \cite{dt}, we arrive to the conclusion that the addition of the $\log \nu^2$ term to the one-loop amplitude done in this work cancels the IR singularities in a similar way to including the effect of soft gravitons with a momentum up to a scale $\mu$. Thus, this scale plays the role 
of the scale separating soft from hard gravitons or, in other words, the scale defining what is considered a hard enough graviton to be detected. In any event, it is important to remember in the following that, due to the different approximations considered, $\mu$ has always to be taken much smaller than the energy of the scattering processes, i.e.
$\mu^2 << s$. In addition, as we are considering EH gravity as an effective low-energy theory we have the additional constraint $s< M_P^2$. In summary, the range of applicability of our results will be $\mu^2<<s< M_P^2$.

\section{Unitarization and the IAM method}

In order to solve the unitarity problem, at least partially, one could try to implement any of the well known available unitarization methods like K-matrix, N/D or the Inverse Amplitude Method (IAM) (see \cite{Delgado:2015kxa} for a comparison among these methods for $W_LW_L$ scattering in the context of a strongly interacting symmetry breaking sector of the Standard Model \cite{DHD}). From all of them, the IAM method \cite{Truong:1988zp} seems to be the more appropriate for the kind of computation we are considering here where we have an expansion in $s$ powers, good analytical properties and perturbative elastic unitarity. When this is the case, the IAM produces unitary amplitudes $a_J^{IAM}(s)$ with the right analytical structure (RC and LC in the first Riemann  sheet) and with the correct low-energy behavior. The IAM method has also proved to be very efficient describing dynamical resonances in hadron physics  in the context of unitarized Chiral Perturbation Theory \cite{Dobado:1992ha}.

Starting from the perturbative 
first two terms (LO and NLO), the IAM partial waves are defined for our helicity \pppp\ case as:
\begin{equation}\label{eq:IAM}
    a_J^{IAM}(s)= a_J^{(0)}(s)\frac{a_J^{(0)}(s)}{a_J^{(0)}(s)-a_J^{(1)}(s)}.
\end{equation}
From this definition it is straightforward to show exact elastic unitarity:
\begin{equation}
\Imag    a_J^{IAM}(s)= \left\lvert a_J^{IAM}(s) \right\rvert ^2
\end{equation}
on the RC (physical region where $s=E_{CM}^2+i0$) provided
\begin{equation}
\Imag    a_J^{(1)}(s)= \left( a_J^{(0)}(s) \right)^2 
\end{equation}
also on the RC. At the same time, at low energies $s\ll M_P^2$ we have 
the expansion:
\begin{equation}
    a_J^{IAM}(s)= a_J^{(0)}(s)+a_J^{(1)}(s)+...
\end{equation}
Therefore (neglecting $O(\mu^2/s)$ corrections) the IAM partial wave defined above has all the expected properties of the \pppp\ (and \mmmm) wave.

In addition, because of the particular structure of the IAM amplitude, the unitarized partial waves can potentially show poles on the complex plane $s$. Thus, according to general $S$-matrix theory \cite{Dispers2}, if these poles appear in the second (unphysical) Riemann sheet under the RC, they have the natural interpretation of dynamical resonances produced by the graviton-graviton interaction. In this case, their location $s_0$ on the complex plane will define the resonance mass $M_R$ and width $\Gamma_R$ as:
\begin{equation}\label{eq:masswidth}
    s_0= M_R^2- i M_R \Gamma_R.
\end{equation}
On the other hand, if the poles appear on the first (physical) Riemann sheet, they are ghosts (spurious states) and must be interpreted as artifacts of the different approximations considered.

It is this last particular property of the IAM partial waves the one that we want to exploit more in this work. In the next section we will perform an analytical continuation of the \pppp\ IAM amplitudes to the second Riemann sheet seeking for poles that could be interpreted as dynamical resonances of graviton-graviton scattering for different $J$ values.

It is clear from the outset that the interpretation of the results will be made difficult by the presence of the IR regulator (our scale $\mu$). According to our previous discussion our infrared regulator $\mu$ is the scale separating soft from hard gravitons (or equivalently the resolution of the graviton detector). Therefore our results will be dependent on this parameter. However, as we have to fulfill the constraint $\mu^2\ll s < M_P^2$ as discussed above, we will be able to establish several relevant general $\mu$ independent conclusions concerning the possibility of graviton scattering resonances.

\section{Numerical results}
In this section and in the figures we will follow the usual convention to label the complex plane quadrants: I (real and imaginary parts both positive), II (real part negative, imaginary one positive), III (real and imaginary parts both negative) and IV (real part positive, imaginary one negative). Also $s$ is given in $M_P^2$ units in this section.

The physical partial wave amplitudes $a_J(s)$ are evaluated above the real axis. That is, $a_J(s) = a_J(E^2_{CM}+i 0)$. However, according to $S$-matrix theory, the partial waves $a(s)$ are analytic and can be analytically continued to the complex plane~\cite{Dispers2}.

Since the partial wave amplitudes $a_J(s)$ that we are considering are compositions of polynomials, rational functions and logarithms, the analytical continuation of the logarithms is the only one that is non trivial. On the first Riemann sheet,
\begin{align*}
   \log^I(s) &= \log\lvert s\rvert + i\arg(s) \\
   \log^I(-s) &= \log\lvert s\rvert + i\arg(-s),
\end{align*}
where $\arg(s)$ has a let cut along the negative real axis. Due to the Schwartz reflection principle~\cite{Dispers2} we have; $a_J(s+i\epsilon)-a_J(s-i\epsilon)=2i\Imag a_J(s+i\epsilon)$. Thus the first Riemann sheet has a cut over the real axis, but its value over such axis (quadrants I and II) is an analytical continuation from $a_J(E_{CM}^2+i0)$. Also the quadrants III and IV are a mirror reflection of quadrants II and I.

On the other hand, the second Riemann sheet is the analytic continuation to quadrant  IV crossing the positive part of the real axis. Quadrants I and II are the same as in the first Riemann sheet. All the figures on the complex plane will refer to the {\it second} Riemann sheet.

In our analysis, we are using the Wolfram Mathematica framework. By default, this package defines the complex logarithmic function $\log s$ having a discontinuity over the negative real axis. Furthermore, the function call $\arg s$, that returns the argument of the complex number $s$, is defined with a discontinuity over the negative real axis, and $\arg s\in (-\pi,\pi]$. With this in mind, we can define the second Riemann sheet relevant logarithm as:
\begin{equation}
   \log^{II} (-s) = \log_M\lvert s\rvert + i\left(\arg_M s - \pi\right),
\end{equation}
where $\log_M$ and $\arg_M$ are the ones defined on Mathematica. 

\begin{figure*}[p]
    \centering
    \includegraphics[width=.8\textwidth]{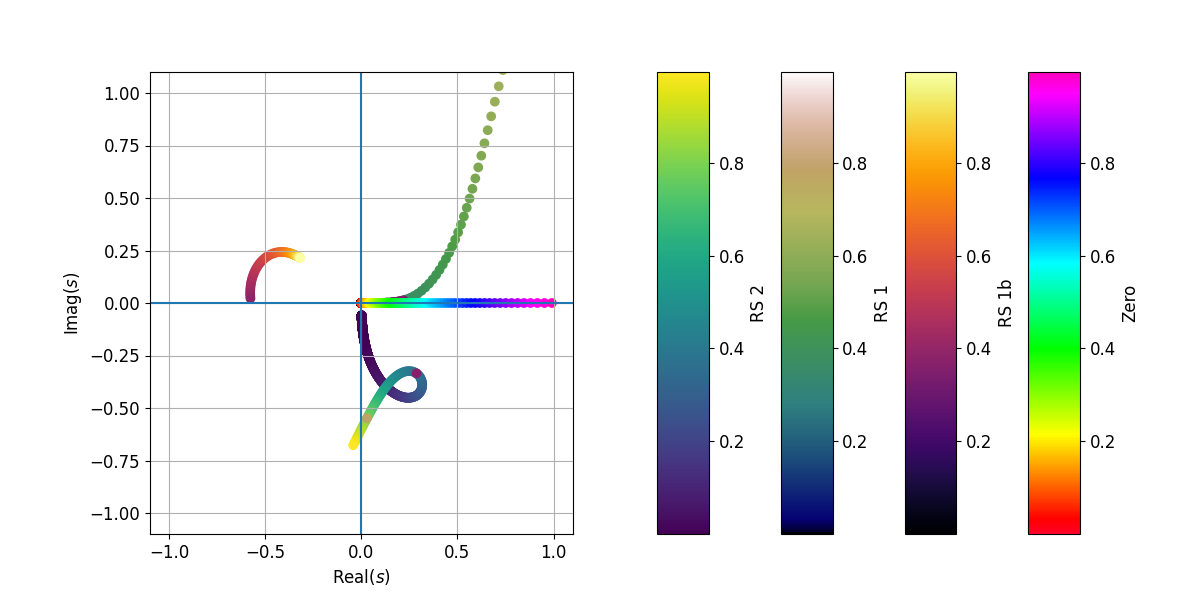}\\
    \includegraphics[width=.8\textwidth]{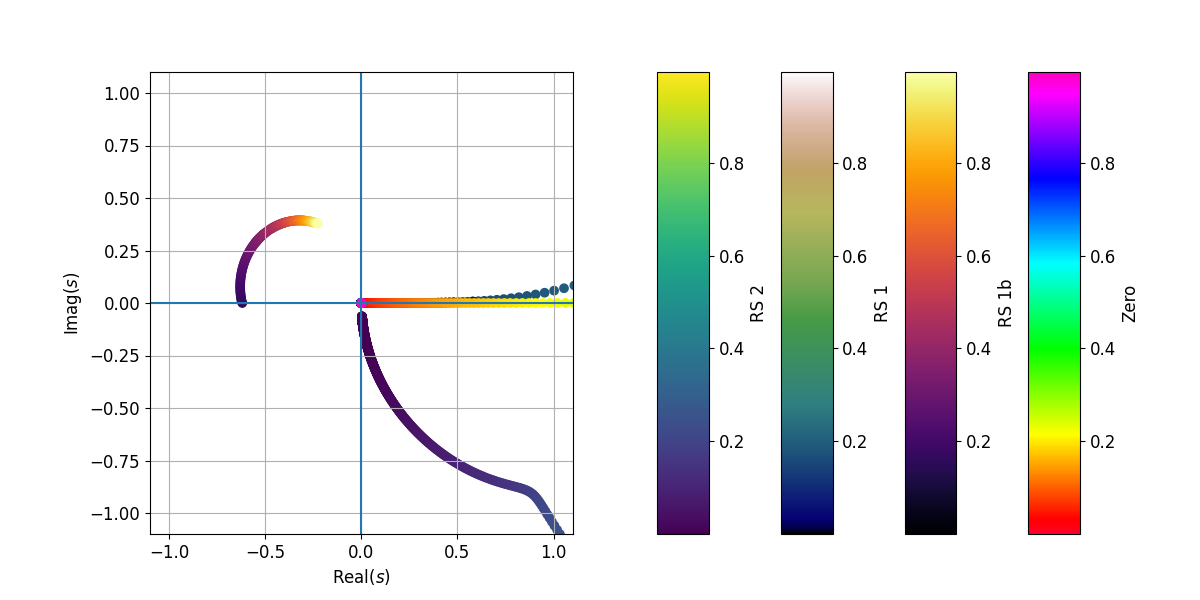}\\
    \includegraphics[width=.8\textwidth]{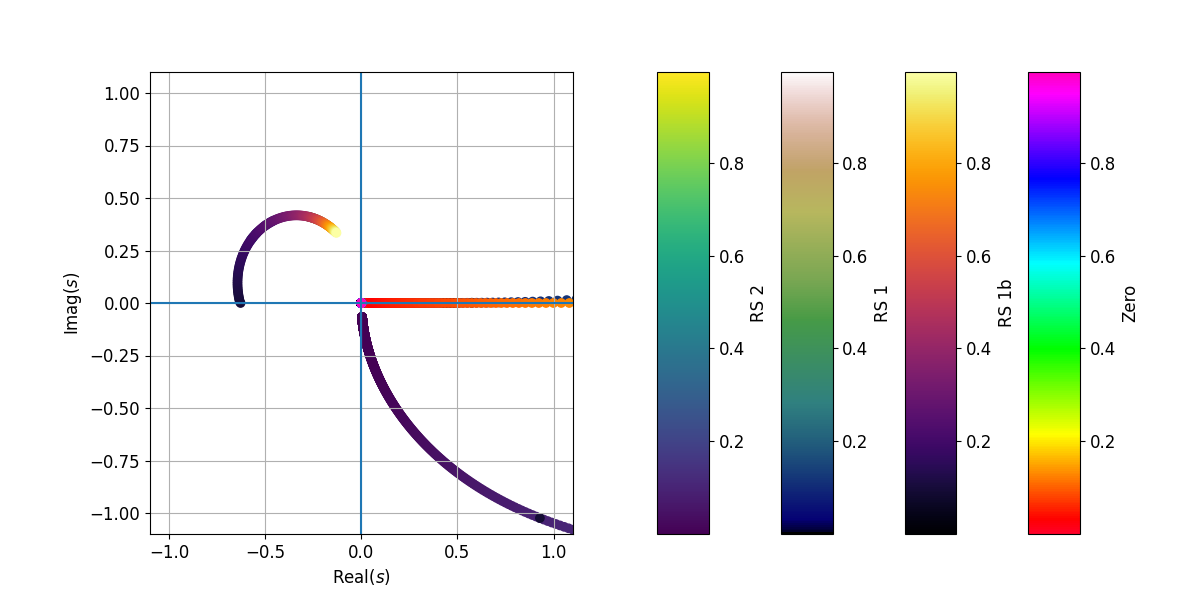}\\
    \caption{From top to bottom, pole positions of $J=0$, $J=2$ and $J=4$ partial waves. The color scales stand for $\mu/M_P$. The \texttt{Zero} points refer to a zero in the numerator of the Inverse Amplitude Method (mostly over the positive real axis); \texttt{RS~1}, \texttt{RS~1b} and \texttt{RS~2}, to poles on the quadrants I, II and IV; and \texttt{Zero}, to zeros on the quadrant I. Color in online version.}
    \label{fig:poleposition}
\end{figure*}
\begin{figure*}
    \centering
    \null\hfill
    \includegraphics[width=.42\textwidth]{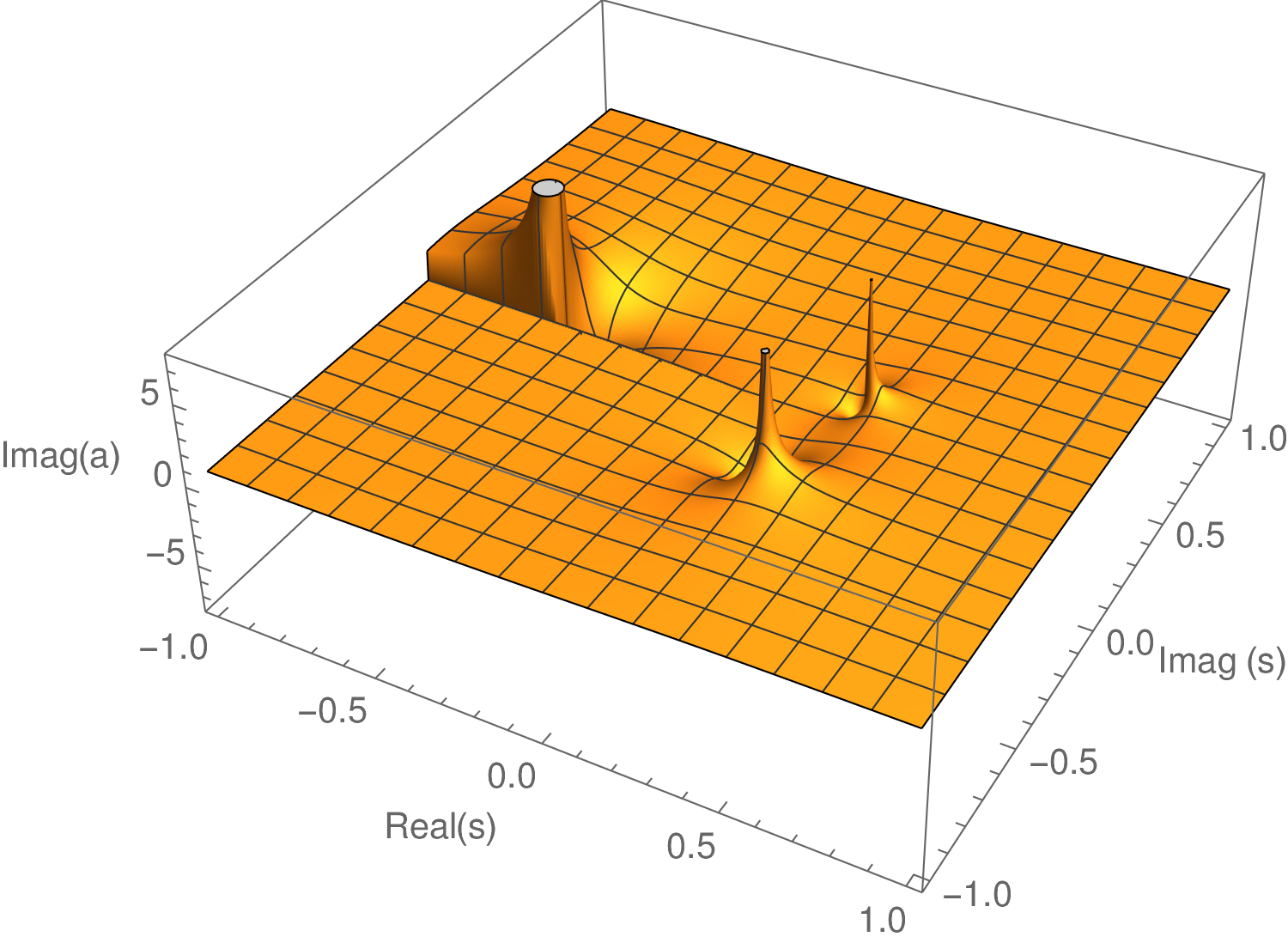} \hfill
    \includegraphics[width=.42\textwidth]{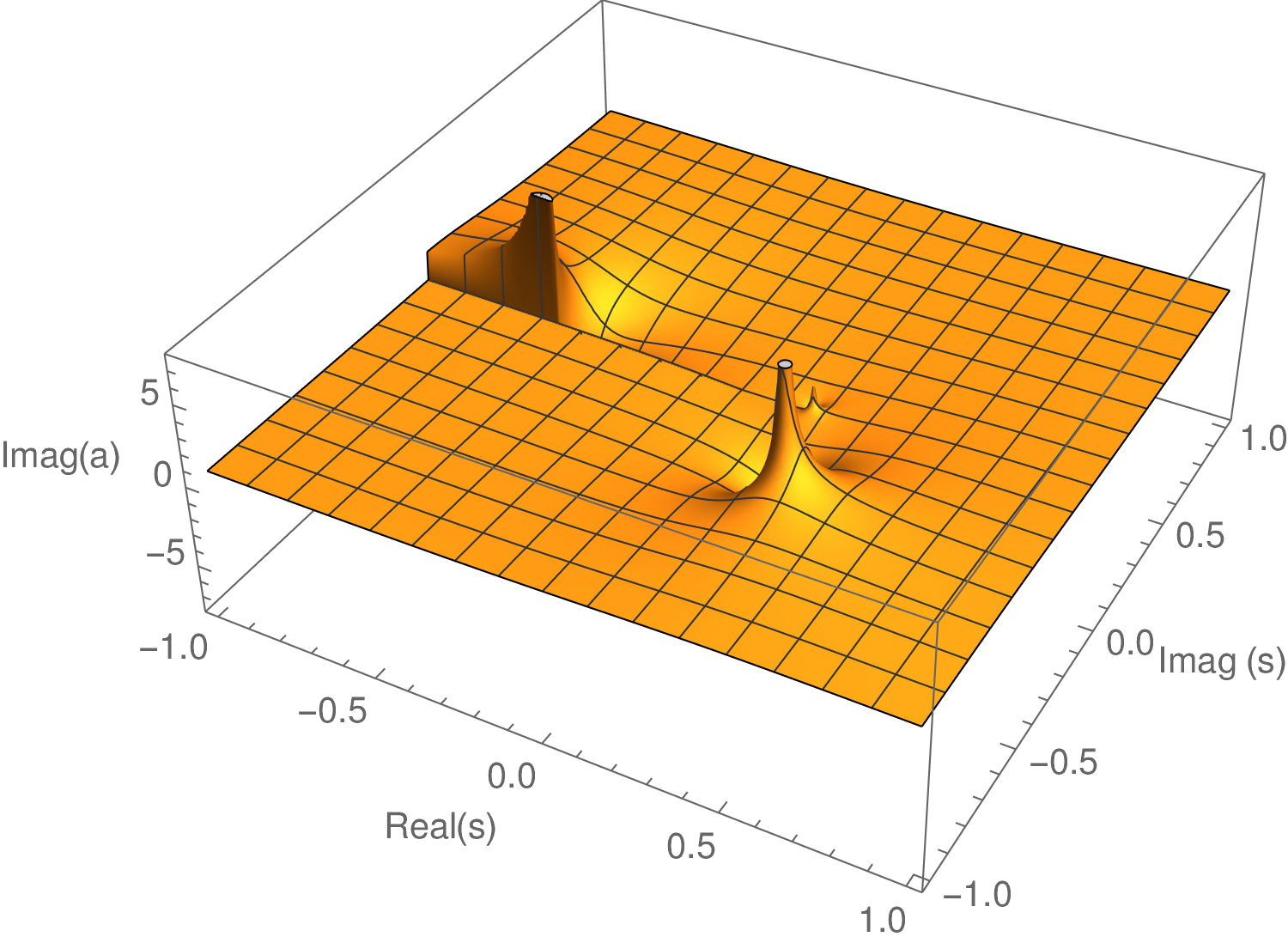} \hfill\null\\
    \null\hfill
    \includegraphics[width=.42\textwidth]{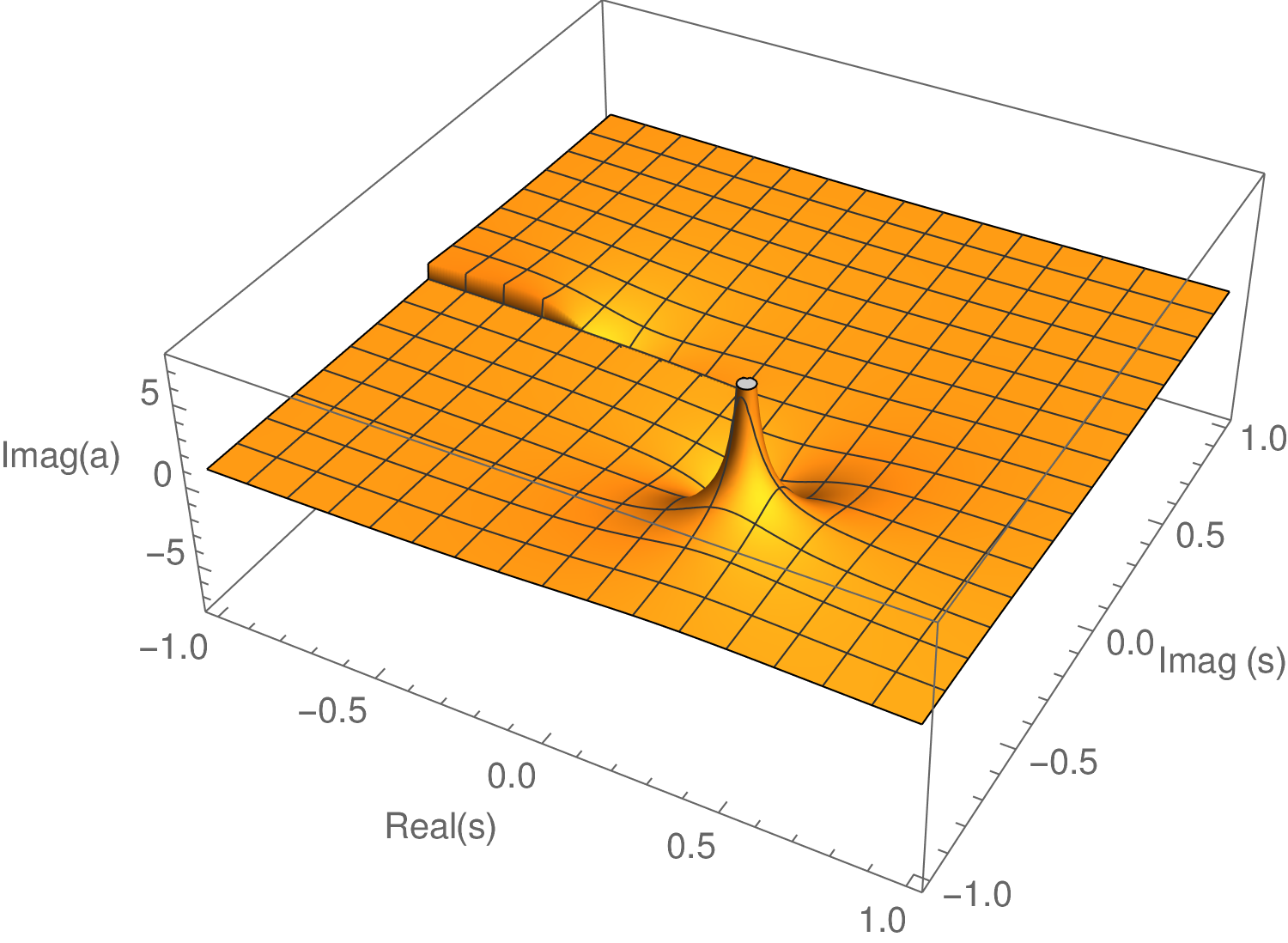} \hfill
    \includegraphics[width=.42\textwidth]{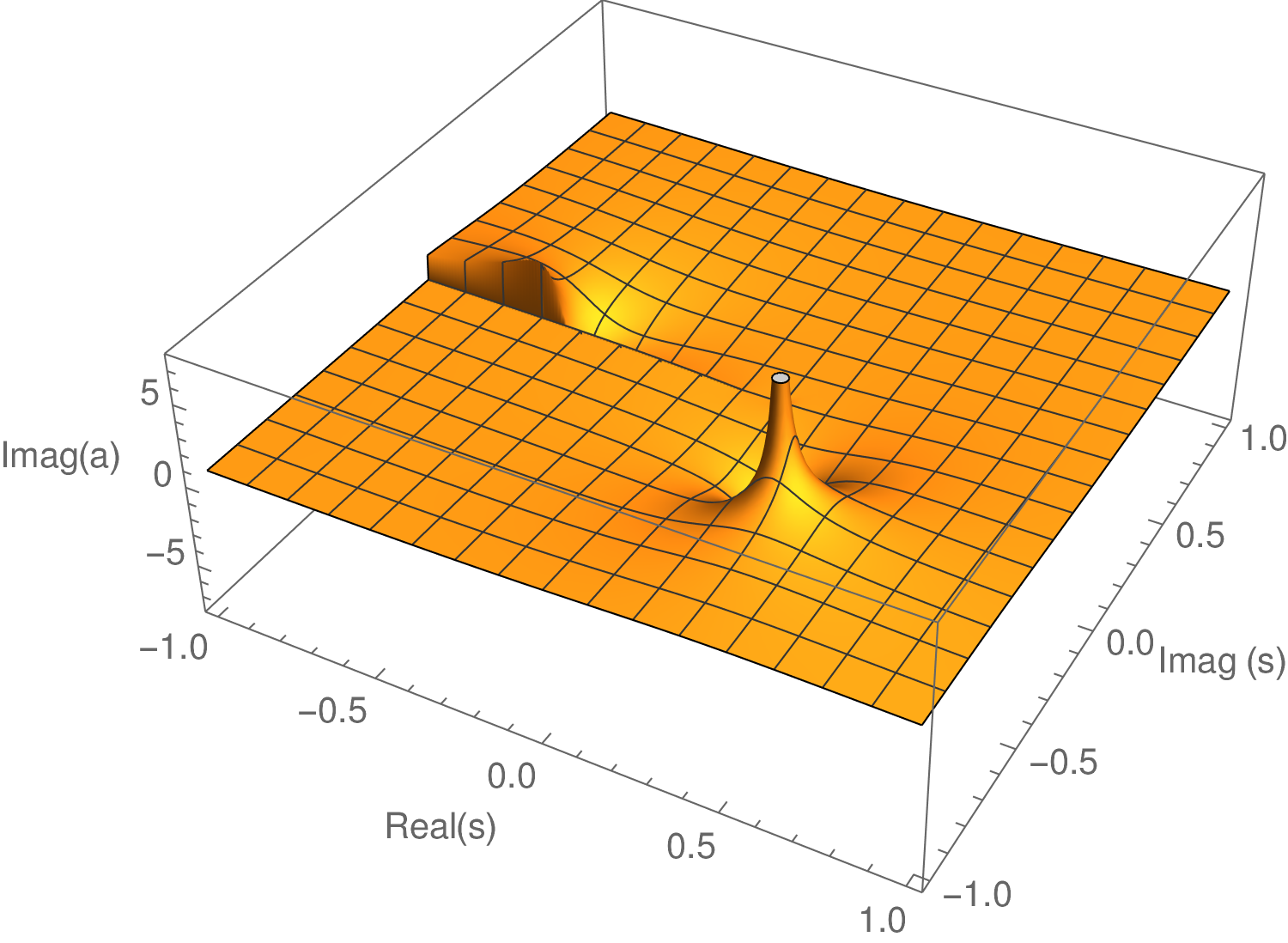} \hfill\null\\
    \caption{From top left, clockwise: plots of $\Imag a_0^{IAM}$ for $\mu/M_P=0.40,\,0.35,\,0.30,\,0.20$. Notice the disappearance of the poles on the first Riemann sheet (quadrants I and II).}
    \label{fig:plots_J0}
\end{figure*}

\begin{figure*}
    \centering
    \null\hfill
    \includegraphics[width=.42\textwidth]{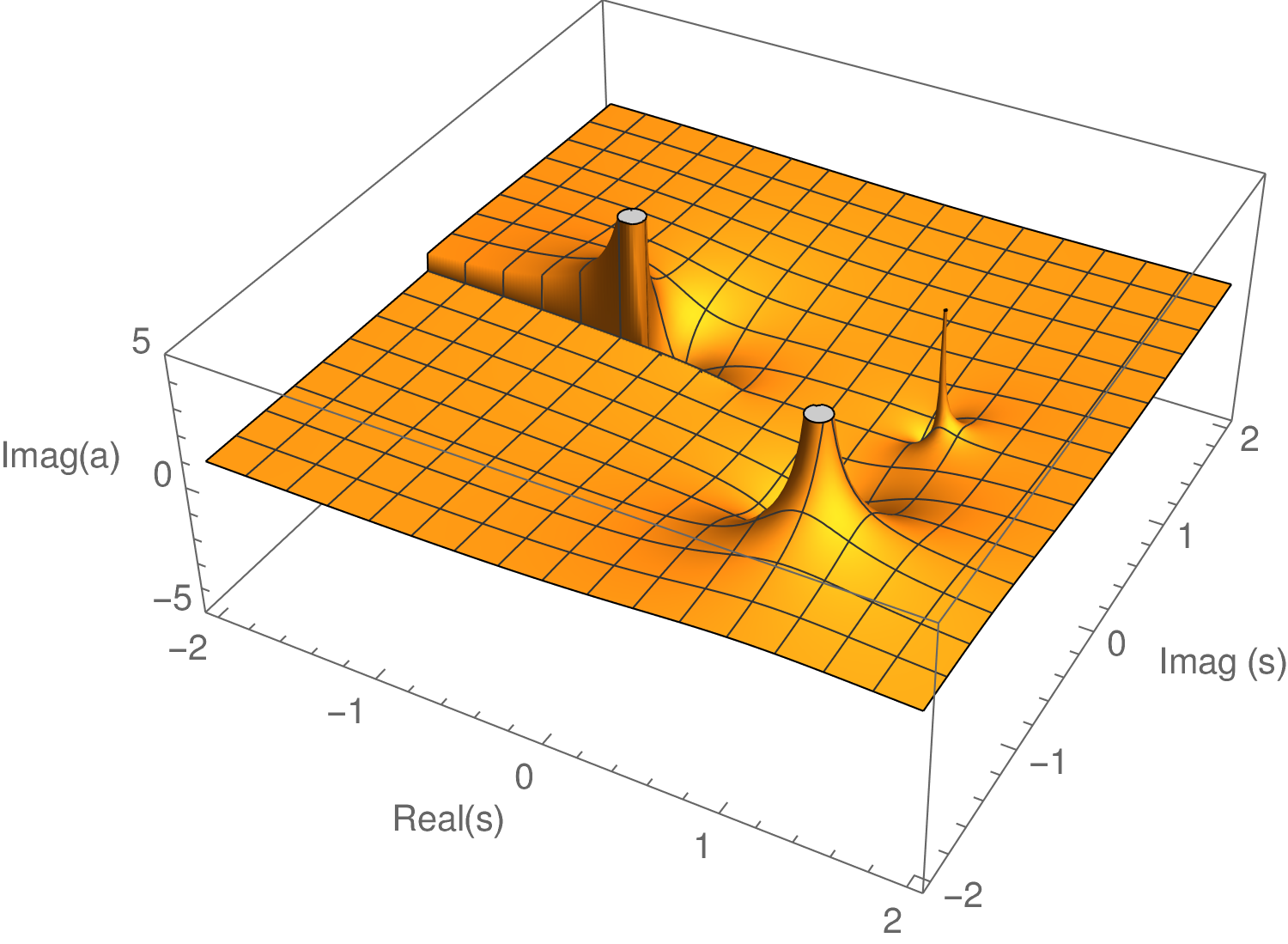} \hfill
    \includegraphics[width=.42\textwidth]{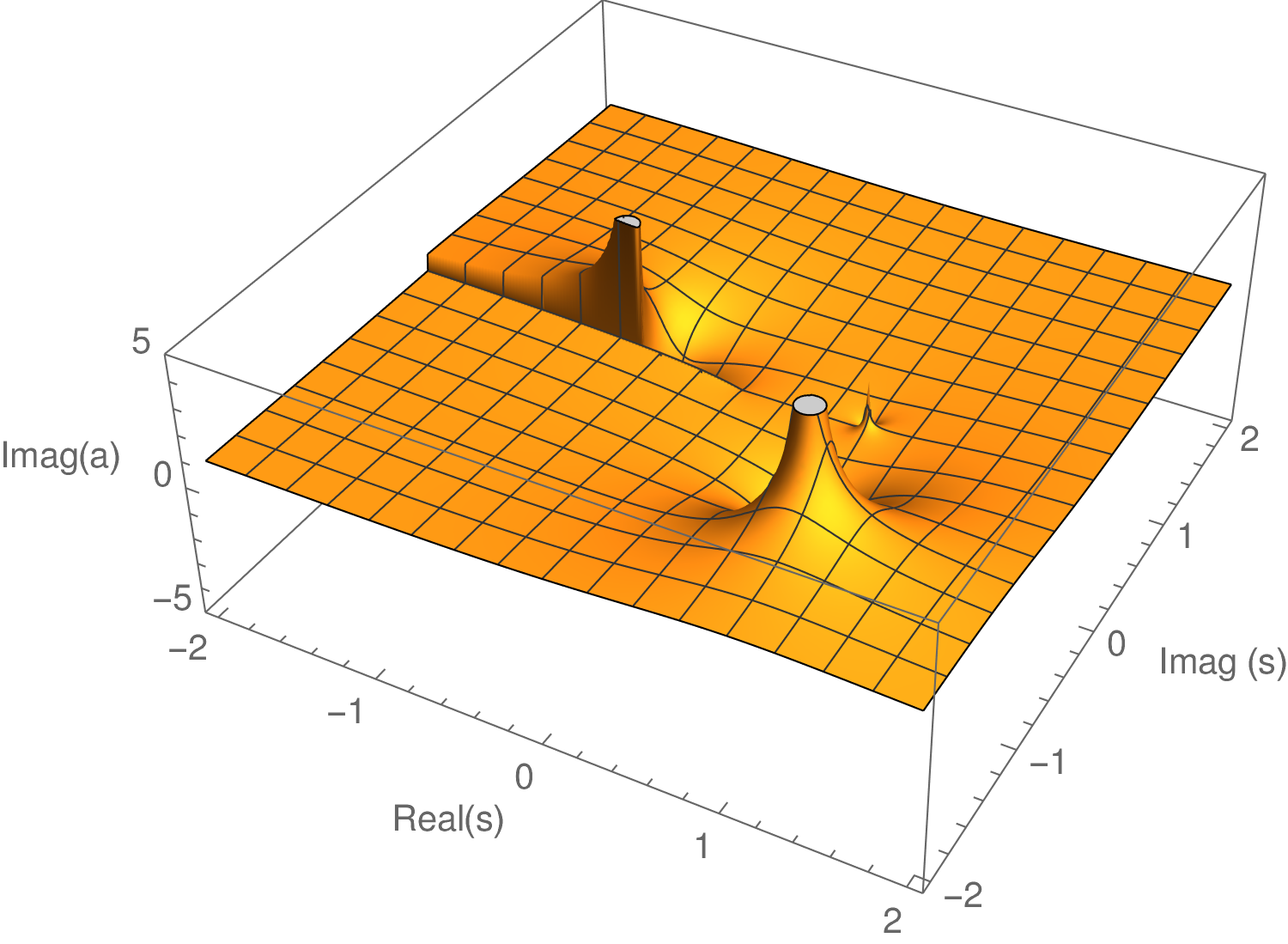} \hfill\null\\
    \null\hfill
    \includegraphics[width=.42\textwidth]{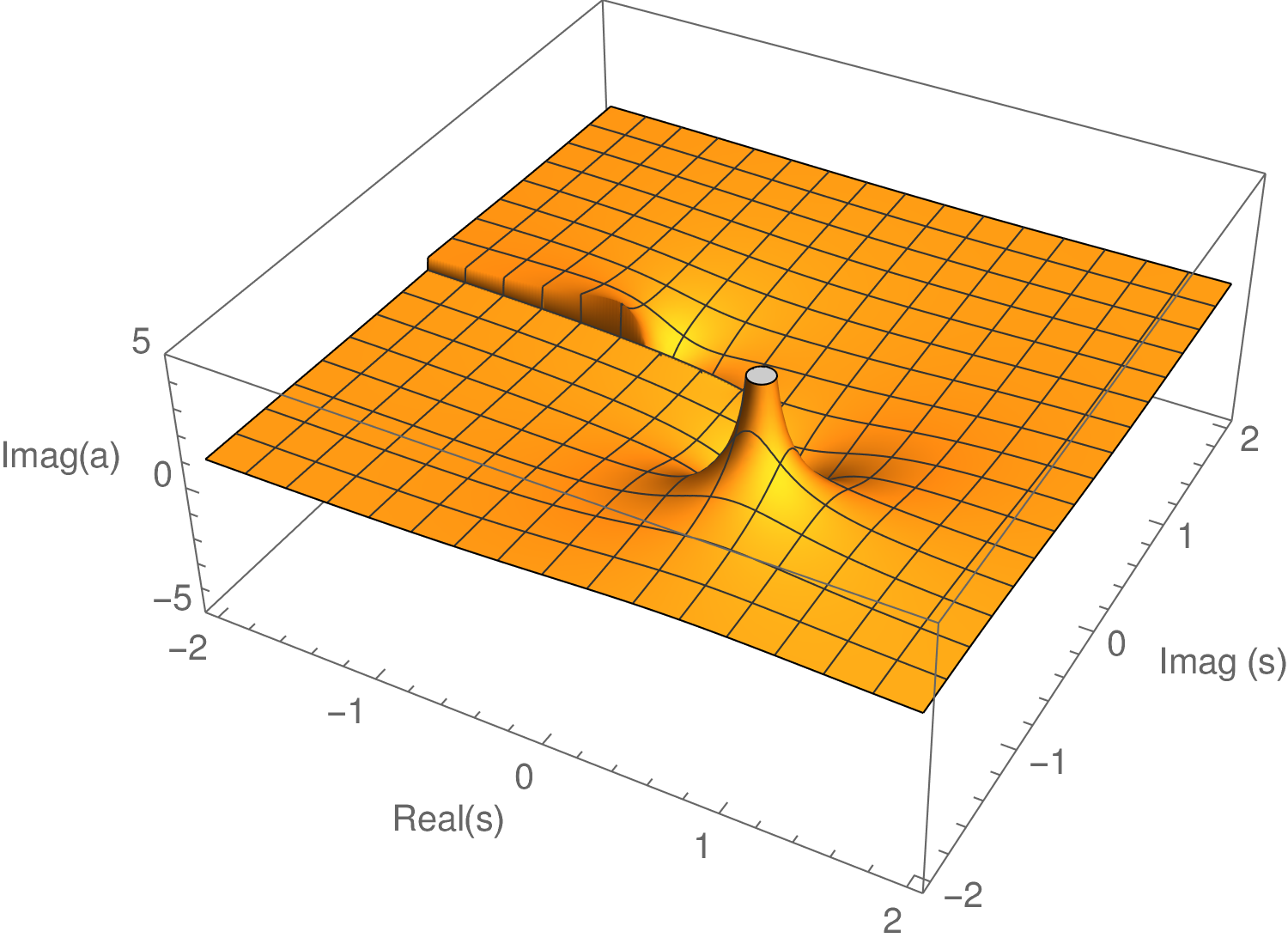} \hfill
    \includegraphics[width=.42\textwidth]{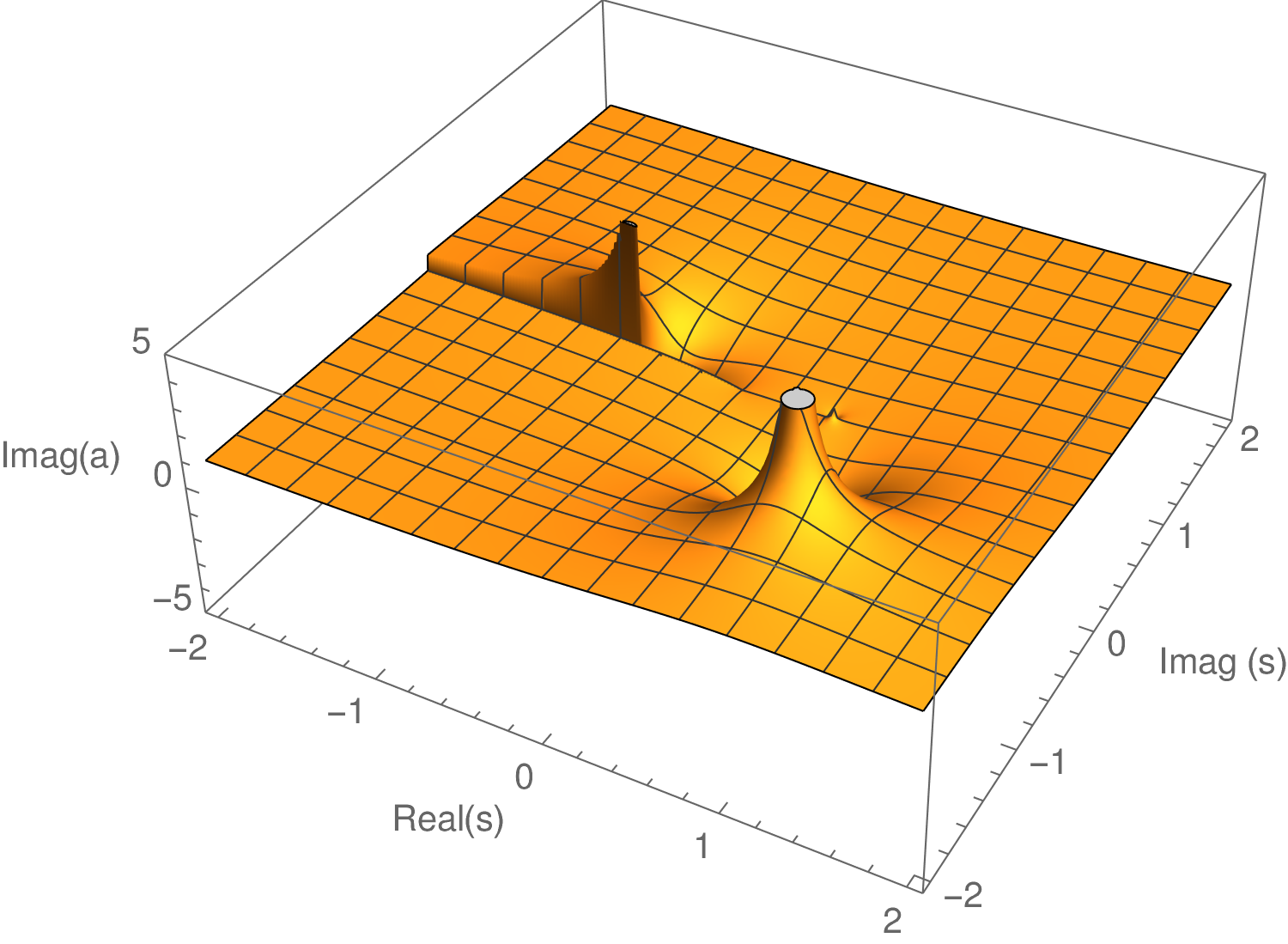} \hfill\null\\
    \caption{From top left, clockwise: plots of $\Imag a_2^{IAM}$ for $\mu/M_P=0.20,\,0.17,\,0.15,\,0.10$. Notice the disappearance of the poles on the first Riemann sheet (quadrants I and II).}
    \label{fig:plots_J2}
\end{figure*}

In fig.~\ref{fig:poleposition} we study the pole positions for different values of the rate $\mu/M_P$. Poles are searched on all the quadrants of the second Riemann sheet\footnote{Quadrants I and II are the same than in the first Riemann sheet~\cite{Dispers2}.} as zeros on the denominator of eq.~\ref{eq:IAM}. However, it could happen that both, the numerator and the denominator of eq.~\ref{eq:IAM}, cancel at some point while the actual function has no pole there. Hence, we also look for zeros on the numerator of eq.~\ref{eq:IAM}.
Indeed, there is a pole located over the real axis on the I quadrant (both real and imaginary parts of $s$ positive), but tends to cancel with a zero on the numerator of eq.~\ref{eq:IAM} on the limit $\mu\ll M_P$. This can be also seen on figs.~\ref{fig:plots_J0} and~\ref{fig:plots_J2}. For $\mu/M_P<0.10$, the pole on the first quadrant has disappeared from the $J=0$ and $J=2$ plots. A similar result is obtained for $J=4$ and $\mu/M_P < 0.05$. As shown on fig.~\ref{fig:poleposition}, there is also a pole on the II quadrant (negative real part, positive imaginary one). However, this pole vanishes for relatively high values of $\mu/M_P<0.37$ ($J=0$), $\mu/M_P<0.15$ ($J=2$) and $\mu/M_P<0.092$ ($J=4$). The disappearance of this pole can also be seen on figs.~\ref{fig:plots_J0} and~\ref{fig:plots_J2}, where the pole above the RC disappears when $\mu/M_P < 0.1$.
Finally, there is a pole on quadrant IV that can be seen on fig.~\ref{fig:poleposition} (from now on, {\it pole on the second Riemann sheet}, since it does not appear on the first Riemann sheet). This pole on the second Riemann sheet could be a resonance, but tends to the origin for sufficiently low values of $\mu\ll M_P$. 

\begin{figure*}
    \centering
    \null\hfill\includegraphics[width=.42\textwidth]{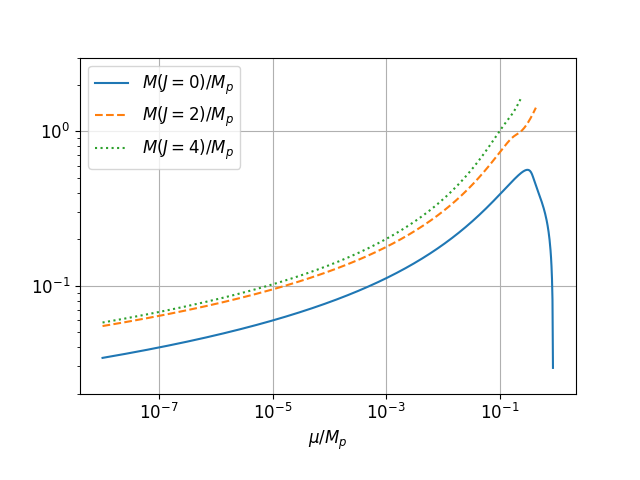} \hfill
    \includegraphics[width=.42\textwidth]{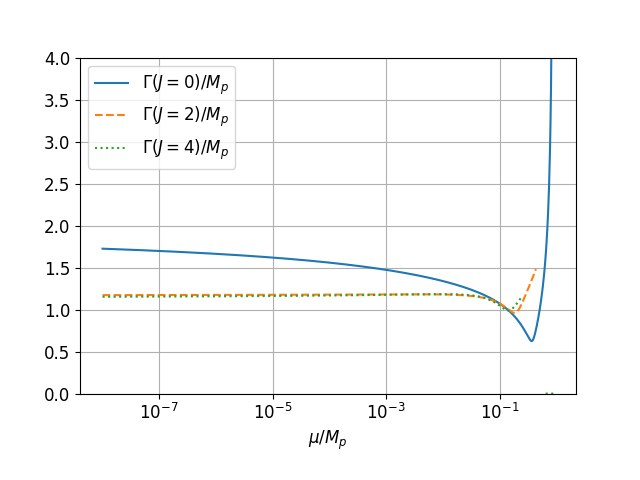} \hfill\null\\
    \caption{From left to right, mass and width of the resonance on the IV quadrant (second Riemann sheet) as a function of $\mu/M_P$. Solid blue line, $J=0$; dashed orange line, $J=2$; and dotted green line, $J=4$. Color in online version.}
    \label{fig:mass_width}
\end{figure*}
\begin{figure*}
    \centering
    \null\hfill
    \includegraphics[width=.42\textwidth]{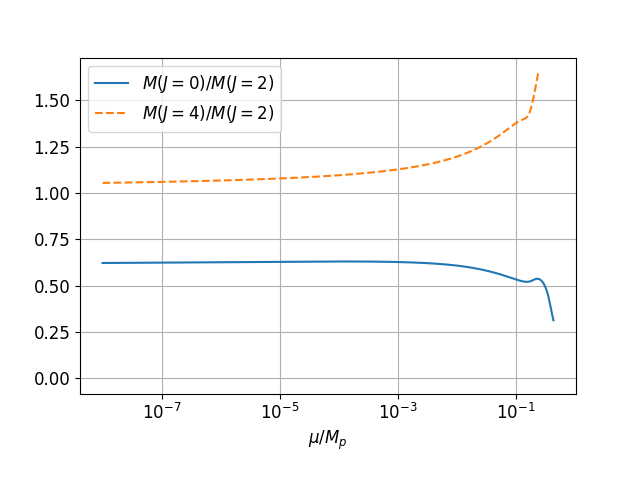} \hfill
    \includegraphics[width=.42\textwidth]{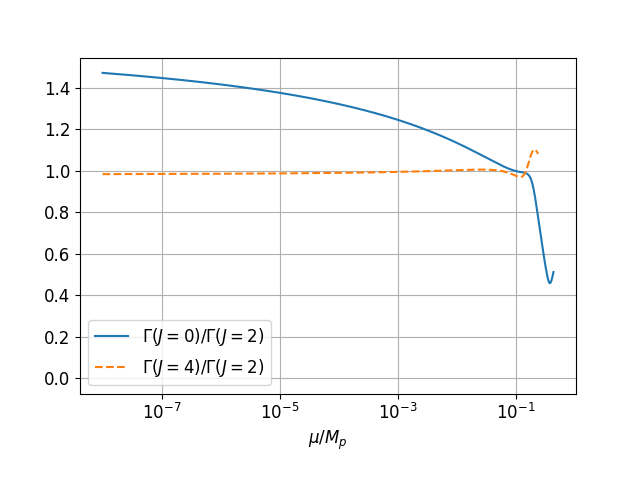} \hfill\null\\
    \caption{From left to right, ratios of mass ($M(J=j)/M(J=2)$) and width ($\Gamma(J=j)/\Gamma(J=2)$) of the resonance on the IV quadrant (second Riemann sheet) as a function of $\mu/M_P$. Solid blue line, $j=0$; and dashed orange line, $j=4$. Color in online version.}
    \label{fig:mass_width_ratio}
\end{figure*}

For studying the behavior of the pole on the second Riemann sheet, the mass and width of the resonance, computed according to eq.~\ref{eq:masswidth}, are plotted on fig.~\ref{fig:mass_width} as a function of $\mu/M_P$. All the masses tend logarithmically to zero when $\mu/M_P\to 0^+$. Concerning the widths, $\Gamma(J=2) \approx \Gamma(J=4)\approx 1.2M_P$ for $\mu < 0.25 M_P$. However, $\Gamma(J=0)$ does not stabilize and, when $\mu\to 0^+$, it grows very slowly. For $\mu=10^{-7}M_P$, $\Gamma(J=0)\approx 1.7M_P$. 

On fig.~\ref{fig:mass_width_ratio}, we plot the ratios $M(J=i)/M(J=2)$ and $\Gamma(J=j)/\Gamma(J=2)$ for $j=0,4$. The ratios $M(J=0)/M(J=2)$ and $M(J=4)/M(J=2)$ stabilize at sufficiently low values of $\mu\ll M_P$ (fig.~\ref{fig:mass_width_ratio}), although all the masses tend to 0 when $\mu\ll M_P$ (fig.~\ref{fig:mass_width}).


\section{Discussion and conclusions}
In this work we have investigated the possibility that dynamical resonances are generated in pure gravity when the Einstein-Hilbert theory 
is interpreted in the context of effective field theory as the low energy description of a more general UV completion.  In low-energy hadron physics, such an approach is able to predict the existence of resonances (e.g. vector mesons) that actually exist in QCD. If one adopts such a view, the main difference between the chiral Lagrangian and the EH theory is the
vanishing of the ${\mathcal O}(p^4)$ coefficients, absent in EH. In strong interactions they are known to be mostly responsible for the presence of vector and scalar 
dynamically generated resonances. But in EH theory, in the pure gravity case, they vanish on-shell.

In the present setting, due to the presence of infrared singularities in the amplitudes,  it is unavoidable to introduce an IR regulator $\mu$. This regulator plays the role of the scale separating soft (not detected) from hard gravitons. In any case, because of the way it is introduced in the computations it is clear that $\mu \ll M_P$.

According to the results of the previous section, there are no causality breaking poles on the first Riemann sheet for low values of $\mu/M_P<0.092$. There is a resonance on the second Riemann sheet, but its mass tends logarithmically to zero when $\mu/M_P\to 0$. Numerical instabilities prevent us from going to extremely low values for this ratio, but we find no evidence whatsoever that, in the limit where $\mu \ll M_P$,  
that is, in the physical region where the EH Lagrangian can be consistently interpreted as an effective theory, any resonance is present. In fact, as discussed in Appendix I, we find a scalar resonance similar to that found
in~\cite{Blas:2020och} (graviball) at $\mu/M_P=0.176$, but this value is well above the limit of applicability 
of the unitarized amplitudes  based in the absence of ghosts in the first Riemann sheet which is $\mu/M_P<0.092$.

In order to find a different result one should probably consider higher dimensional operators with non-zero coefficients. That means moving away from EH theory or, perhaps, consider the effect of matter fields coupled to gravity. As we have mentioned above, although in unitarized Chiral Lagrangian for low-energy hadron physics or in the effective theory treatment of the symmetry  breaking sector of the Standard Model, the ${\mathcal O}(p^4)$ terms may lead to resonances, this possibility is prohibited in gravity unless one wants to circumvent the symmetry principles behind general relativity.

In summary, we do not find strong enough evidence for the existence of a graviball in pure gravity, at least at sub-planckian scales, in EH theory understood as an effective low-energy theory for gravitation.

\section*{Acknowledgements}
The authors thank J.A. Oller for reading the manuscript and useful comments. This research is partly supported by the Ministerio de Ciencia e Innovación under research grants PID2019-108655GB-I00, PID2019-105614GB-C21, PID2021-124473NB-I00 (R.~L.~Delgado) and the “Unit of Excellence Mar\'ia de Maeztu 2020-2023” award to the Institute of Cosmos Sciences (CEX2019-000918-M),
and the grant 2017-SGR-929 from Generalitat de Catalunya.
R.~L.~Delgado was also financially supported by the Ram{\'o}n Areces Foundation post-doctoral fellowship, the Istituto Nazionale di Fisica Nucleare (INFN) post-doctoral fellowship AAOODGF-2019-0000329.

\appendix

\section*{Appendix I: Unitarization of the tree level amplitude}
If it were not for the one-loop elastic graviton scattering computation by Dunbar and Norridge \cite{Dunbar:1994bn}, one could consider the possibility of unitarizing the tree level amplitude. For example, for the case \pppp, $J=0$ tree level partial wave $a_0^{(0)}(s)$, one could introduce the unitarized amplitude $a_0(s)$ defined by the simple formula:
\begin{equation}
a_0(s)=\cfrac{a_0^{(0)}(s)}{1+\cfrac{a_0^{(0)}(s)}{\pi}\log \cfrac{-s}{\Lambda^2}}.
\end{equation}
This partial wave shows an unitary RC and in the physical region fulfills elastic unitarity:
\begin{equation}
\Imag a_0(s)=\lvert a_0(s) \rvert ^2.
\end{equation}
However, this unitarization method introduces an arbitrary (typically UV) scale $\Lambda$ which is an artifact of this particular unitarization scheme. In any case, a proper definition of the tree level partial wave amplitude $a_0^{(0)}(s)$, requires the introduction of a genuine IR regulator (called $\mu$ in this work).

The $a_0(s)$ partial wave above can be extended to the second Riemann sheet by
using \cite{Dispers2}:
\begin{equation}
a_0^{II}(s)=\cfrac{a_0(s)}{1-2ia_0(s)}
\end{equation}
and one can seek for poles of this second Riemann sheet in the IV quadrant fulfilling:
\begin{equation}
a_0(s_0)+\cfrac{i}{2}=0
\end{equation}
from which one gets the secular equation:
\begin{equation}
1+\cfrac{a_0^{(0)}(s_0)}{\pi}\log\cfrac{-s_0}{\Lambda^2}-2i a_0^{(0)}(s_0)=0. 
\end{equation}
In~\cite{Blas:2020och} the authors used this method, adding some assumptions, to find a pole at $s_0 =(0.07-i~ 0.20) \Lambda^2$ in the $J=0$ channel
which they claim is a pure gravitational resonance (graviball). However this claim is questionable for at least two reasons.

First, the position of the pole depends on the UV arbitrary scale $\Lambda$ which is an artifact of the unitarization method used, since pure elastic graviton scattering is UV finite up to one-loop. Second, independently of the $\Lambda$ value, the width associated to the pole is so large compared with its mass that hardly could it be considered a physical state in the usual sense.

\begin{figure*}
    \centering
    \null\hfill
    \includegraphics[width=.45\textwidth]{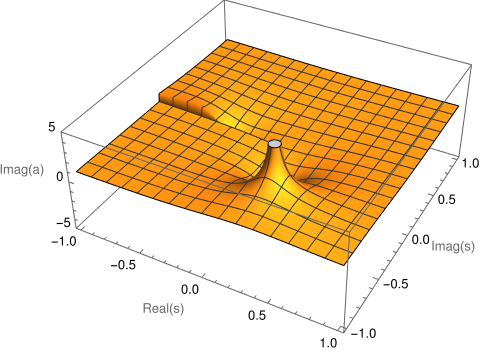} \hfill\null\\
    \includegraphics[width=.40\textwidth]{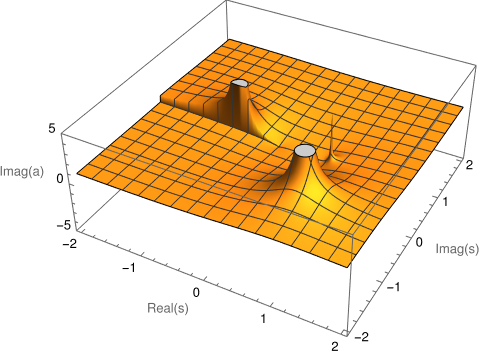} \hfill
    \includegraphics[width=.40\textwidth]{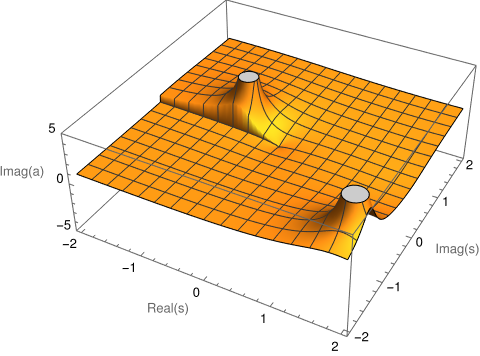} \hfill\null\\
    \caption{From top, anticlockwise: plots of $\Imag a_0^{IAM}$ for $J=0$, $J=2$ and $J=4$; $\mu/M_P=0.176$.}
    \label{fig:plots_Oller}
\end{figure*}
In this work we are using a more robust unitarization method which does not require the introduction of arbitrary new UV scales but only the IR regulator $\mu$ already present in the perturbative computations before unitarization. Hence, it seems interesting to check if this pole in ref.~\cite{Blas:2020och} can be reproduced by using the method introduced in this work. Therefore we have looked for the closest pole we can find in our computations by minimizing the distant to their $s_0$ on the complex $s$ plane by varying our $\mu$ IR regulator. Thus we have found a pole at $s'_0=(0.23-i0.45)M_P^2$ corresponding to $\mu/M_P=0.176$. However, for this $\mu$ value ghosts are present in the $J=2$ and $J=4$ channels, as it can be checked on fig.~\ref{fig:plots_Oller}. In addition  this $\mu/M_P$ value requires to define gravitons as those gravity excitations with a momentum larger than roughly a fifth of the Plack mass scale.

Also in \cite{Blas:2020och} the authors claim the graviball is 
similar to the $\sigma$ particle in the context of low-energy hadron physics \cite{Pelaez:2015qba}. It is true that this very broad $\sigma$ resonance can be obtained by using the unitarization method described in this appendix from the tree level Chiral Lagrangian result for the $J=0$ channel for $\Lambda$ of the order of 1~GeV. However, this is not the case of the $\rho$ resonance in the $J=1$ channel, which requires the introduction of additional information in the form of subtraction constants or chiral parameters like $l_1$ and $l_2$ (see for example \cite{Dobado:1992ha}). 

\section*{Appendix II: One loop contribution to the partial waves}
The tree level partial waves $a_J^{(0)}(s)$ were given in section~\ref{sec:pertunit}. The expressions for the $J=0,\,2,\,4$ one loop $a_J^{(1)}(s)$ partial waves are:
\begin{widetext}
\ban   
a_{0++++}^{(1)}(s) &=& \frac{s^2}{4\pi M_P^4}
  \left[\log^2\frac{s}{\mu^2}F\left(\frac{s}{\mu^2}\right) 
  - \frac{\pi^2}{3}\log\frac{s}{\mu^2} + 2\zeta(3)
  + \frac{173\pi^2}{126} + \frac{1447}{720}\right] \nonumber \\
a_{2++++}^{(1)}(s) &=& \frac{s^2}{4\pi M_P^4}
  \left[\left(\log^2\frac{s}{\mu^2} - 6\log\frac{s}{\mu^2} + 9\right)F\left(\frac{s}{\mu^2}\right) 
  +\left(3-\frac{\pi^2}{3}\right)\log\frac{s}{\mu^2} \right.\nonumber\\
  &&\left. + \pi^2-12 + 2\zeta (3)
  + \frac{173\pi^2}{630} -\frac{43}{2160} \right] \nonumber\\
a_{4++++}^{(1)}(s) &=& \frac{s^2}{4\pi M_P^4}
  \left[\left(\log^2\frac{s}{\mu^2} - \frac{25}{3}\log\frac{s}{\mu^2} + \frac{625}{36}\right)F\left(\frac{s}{\mu^2}\right) 
  +\left(\frac{115}{36} - \frac{\pi^2}{3}\right)\log\frac{s}{\mu^2} \right.\nonumber\\
  &&\left. + \frac{25}{18}\pi^2 - \frac{865}{54} + 2\zeta(3) 
  -\frac{37\pi^2}{4158} + \frac{4139}{3240} \right]\nonumber\\
a_{0+-+-}^{(1)}(s) &=& a_{2+-+-}^{(1)}(s) = 0 \nonumber\\
a_{4+-+-}^{(1)}(s) &=& \frac{s^2}{4\pi M_P^4}
  \left[\left(\frac{1}{2}\log^2\frac{s}{\mu^2} - \frac{363}{140}\log\frac{s}{\mu^2} + \frac{1566947}{352800}\right) 
F\left(\frac{s}{\mu^2}\right)  \right.\nonumber\\
  &&\left. +\left(\frac{419017}{352800} - \frac{\pi^2}{6}\right)\log\frac{s}{\mu^2} 
  +\frac{121}{280}\pi^2 
 -\frac{7015147}{1372000}+\zeta(2)
 -\frac{166097}{176400}F\left(\frac{s}{\mu^2}\right) +\frac{80751073}{74088000}
  \right]\nonumber\\
a_{J+--+}^{(1)}(s) &=& a_{J+-+-}^{(1)}(s),\label{(pw1)}
\ean
\end{widetext}
where we have performed the substitution $\eta\to 2\mu^2/s$.


\begin{thebibliography}{9}

\bibitem{EChL} 
S.~Weinberg, Phys.~Rev.~{\bf 166} (1968), 1568; Physica~{\bf A96} (1979), 327;
J.~Gasser and H.~Leutwyler, Ann.~Phys.~{\bf 158} (1984) 142

\bibitem{Donoghueetal} 
J.~F.~Donoghue, Phys.~Rev.~D {\bf 50} (1994), 3874;
{\it Advanced School on Effective Theories},
eds. F.~Cornet and M.J.~Herrero, World Scientific, Singapore 1997:~gr-qc/9512024; Eur.~Phys.~J.~A \textbf{56}, no.3 (2020), 86;
D. Espriu and D. Puigdom\`enech, Acta~Phys.~Polon.~B {\bf 40} (2009) 3409.

\bibitem{tree}
B.~S.~DeWitt, Phys.~Rev.~{\bf 162} (1967), 1239;
F.~Berends and R.~Gastmans, Nucl.~Phys.~B {\bf 88} (1975), 99;
M.~Grisaru, P.~van~Nieuwehuizen and C.~C.~Wu, Phys.~Rev.~D {\bf 12} (1975), 397

\bibitem{tHV}
G.~'t~Hooft and M.~ Veltman,
Ann.~Inst.~H.~Poincar\'e~A {\bf 20} (1974), 69

\bibitem{weinb}
S.~Weinberg,
Phys.~Rev.~{\bf B140} (1965), 516.

\bibitem{Goroff:1985th}
M.~H.~Goroff and A.~Sagnotti,
Nucl.~Phys.~B \textbf{266} (1986), 709-736

\bibitem{Dunbar:1994bn}
D.~C.~Dunbar and P.~S.~Norridge,
Nucl.~Phys.~B \textbf{433} (1995), 181-208

\bibitem{dobetal}
R.~L.~Delgado, A.~Dobado and F.~J.~Llanes-Estrada,
J. Phys. G \textbf{41}, 025002 (2014)


\bibitem{Blas:2020och}
D.~Blas, J.~Martin Camalich and J.~A.~Oller,
Phys.~Lett.~B \textbf{827} (2022), 136991;
JHEP \textbf{08}, 266 (2022);
J.~A.~Oller,
Phys. Lett. B \textbf{835}, 137568 (2022);
EPJ Web Conf. \textbf{274}, 08011 (2022)

\bibitem{dt}
J.~F.~Donoghue and T.~Torma,
Phys.~Rev.~D {\bf 54} (1996), 4963

\bibitem{Dunbar:1995ed}
D.~C.~Dunbar and P.~S.~Norridge,
Class.~Quant.~Grav.~\textbf{14} (1997), 351-365

\bibitem{Delgado:2015kxa}
R.~L.~Delgado, A.~Dobado and F.~J.~Llanes-Estrada,
Phys. Rev. D \textbf{91} (2015) no.7, 075017;
A.~Dobado and D.~Espriu,
Prog. Part. Nucl. Phys. \textbf{115}, 103813 (2020);
R.~Delgado L\'opez,
{\em Study of the electroweak symmetry breaking sector for the LHC},
Springer Theses, Springer, Berlin Germany (2017).

\bibitem{DHD} A.~Dobado and M.~J.~Herrero,
Phys. Lett. B \textbf{228} (1989), 495-502, {\em ibid} 233 (1989) 505;
J.~F.~Donoghue and C.~Ramirez,
Phys. Lett. B \textbf{234} (1990), 361-366

\bibitem{Truong:1988zp} T. N. Truong,
Phys. Rev. Lett. \textbf{61} (1988), 2526;
Phys. Lett. B \textbf{235} (1990), 134-140;
A. Dobado, M. J. Herrero and T. N. Truong,
Phys. Lett. B \textbf{235} (1990), 129.
 
\bibitem{Dobado:1992ha}
A.~Dobado and J.~R.~Pelaez,
Phys. Rev. D \textbf{47} (1993), 4883-4888;
Phys. Rev. D \textbf{56} (1997), 3057-3073.

\bibitem{Dispers2} R.J. Eden, P.V. Landshoff, D.1. Olive and J.C. Polkingorne, {\em The Analytic $S$-matrix}, 
Cambridge University Press (1966); A.O. Barut, {\em The Theory of the Scattering Matrix}, Macmillan. New York (1967); V Gribov, (Prepared by Y. L. Dokshitzer and J. Nyiri)  {\em Strong Interactions of Hadrons at high Energies}, 
Cambridge University Press (2009); Y.V. Novozhilov, {\em  Introduction to Elementary Particle Theory}, Pergamon Press (1975).


\bibitem{Pelaez:2015qba}
J.~R.~Pelaez,
Phys. Rept. \textbf{658} (2016), 1.

\end{thebibliography}
\end{document}